\documentclass[aps,prb,twocolumn,showpacs,amsmath,amssymb,superscriptaddress,floatfix,nofootinbib]{revtex4-2}

\usepackage[usenames,dvipsnames]{color}         
\usepackage{graphics, nicefrac}      
\usepackage[pdftex]{graphicx}      
\usepackage{bm}
\usepackage{natbib}            
\usepackage[colorlinks,plainpages=false,linkcolor=blue,urlcolor=blue,citecolor=blue,pdfpagemode=UseNone,pdfstartview=FitBH]{hyperref}
\usepackage{float}
\usepackage{array,booktabs,calc}
\usepackage{afterpage}

\begin{document}

\title{Coupling of electronic and structural degrees of freedom in vanadate superlattices}

\author{P.~Radhakrishnan}
\affiliation{Max Planck Institute for Solid State Research, Heisenbergstr. 1, 70569 Stuttgart}

\author{B.~Geisler}
\affiliation{Department of Physics and Center for Nanointegration (CENIDE), Universit\"{a}t Duisburg-Essen, Lotharstr. 1, 47057 Duisburg}

\author{K.~F\"{u}rsich}
\affiliation{Max Planck Institute for Solid State Research, Heisenbergstr. 1, 70569 Stuttgart}

\author{D.~Putzky}
\affiliation{Max Planck Institute for Solid State Research, Heisenbergstr. 1, 70569 Stuttgart}

\author{Y.~Wang}
\affiliation{Max Planck Institute for Solid State Research, Heisenbergstr. 1, 70569 Stuttgart}
\affiliation{Center for Microscopy and Analysis, Nanjing University of Aeronautics and Astronautics, Nanjing 210016, P.R. China}

\author{G.~Christiani}
\affiliation{Max Planck Institute for Solid State Research, Heisenbergstr. 1, 70569 Stuttgart}

\author{G.~Logvenov}
\affiliation{Max Planck Institute for Solid State Research, Heisenbergstr. 1, 70569 Stuttgart}

\author{P.~Wochner}
\affiliation{Max Planck Institute for Solid State Research, Heisenbergstr. 1, 70569 Stuttgart}

\author{P. A.~van Aken}
\affiliation{Max Planck Institute for Solid State Research, Heisenbergstr. 1, 70569 Stuttgart}

\author{R.~Pentcheva}
\affiliation{Department of Physics and Center for Nanointegration (CENIDE), Universit\"{a}t Duisburg-Essen, Lotharstr. 1, 47057 Duisburg}

\author{E.~Benckiser}
\email[]{E.Benckiser@fkf.mpg.de}
\affiliation{Max Planck Institute for Solid State Research, Heisenbergstr. 1, 70569 Stuttgart}

\begin{abstract}
Heterostructuring provides different ways to manipulate the orbital degrees of freedom and to tailor orbital occupations in transition metal oxides. However, the reliable prediction of these modifications remains a challenge.
Here, we present a detailed investigation of the relationship between the crystal and electronic structure in YVO$_3$-LaAlO$_3$ superlattices by combining \textit{ab initio} theory, scanning transmission electron microscopy, and x-ray diffraction. Density functional theory simulations including an on-site Coulomb repulsion term, accurately predict the crystal structure and in conjunction with x-ray diffraction, provide an explanation for the lifting of degeneracy of the vanadium $d_{xz}$ and $d_{yz}$ orbitals, that was recently observed in this system.
In addition, we unravel the combined effects of electronic confinement and octahedral connectivity by disentangling their impact from that of epitaxial strain.
Our results demonstrate that the specific orientation of the substrate and the thickness of the YVO$_3$ slabs in the multilayer, can be utilized to reliably engineer orbital polarization.
\end{abstract}
\maketitle

\section{INTRODUCTION}
Epitaxial growth has emerged as an efficient means to tailor the remarkable traits unique to transition metal oxides (TMO) \cite{Rondinelli2011, Ramesh2019, Rondinelli2012}.
The altered crystal structure of an epitaxial film or multilayer can induce sizeable changes in its electronic properties \cite{Lupascu2014, Hirai2015, Mandziak2019} and even lead to emergent phenomena, such as magnetism in non-magnetic materials \cite{Brinkman2007}, superconductivity in insulators \cite{Gariglio2015, Liu2021} and room temperature ferroelectricity \cite{Haeni2004, Yuan2019}.
\begin{figure*}[t]
	\includegraphics[width=\textwidth]{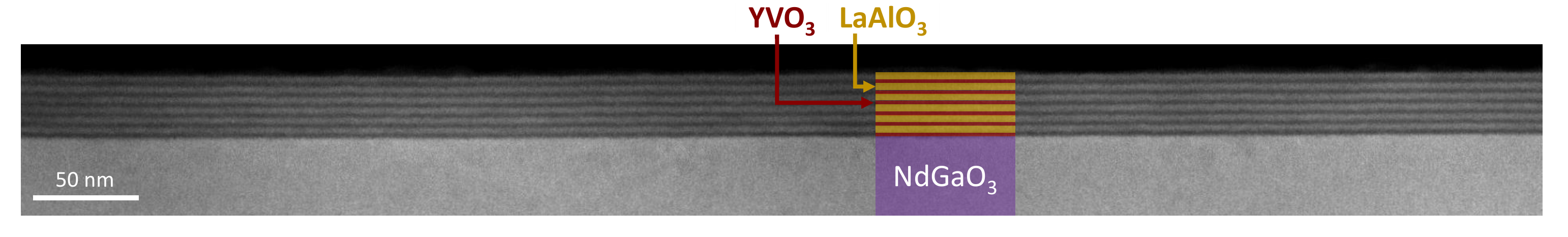}
\caption{Low-magnification annular dark-field STEM image of the YVO/LAO (4/8) \!$\times$\! 6 SL. The absence of defects or stacking faults displays the high quality of the sample, over a large scale up to a micrometer.}
\label{fig0}
\end{figure*}

Modifications of orbital occupations and orbital overlap govern the strength and sign of exchange interactions and therefore determine the magnetic, optical and transport properties of a material \cite{KANAMORI1959,Goodenough1955}.
In strongly correlated electron systems, the orbital occupations at the transition metal site are not only determined by local steric effects such as a specific crystal fields, but are also strongly influenced by the inter-site electron-electron interactions.
Therefore, in TMO heterostructures, disentangling the different contributions that lead to a specific orbital occupation is not trivial, for e.g.\ the influence of bond angles and bond lengths on hopping amplitudes is exactly opposite to one another for a given change in lattice parameter \cite{Wang2019}.
Moreover, octahedral connectivity and symmetry mismatch at the interfaces have an enormous impact on the crystal structure and therefore must also be considered to accurately predict the electronic structure \cite{He2010}.

In this context, simulations in the framework of density functional theory including a Hubbard-$U$ term (DFT$+U$), provide valuable insights.
In particular, these calculations have successfully explained the orbital polarization profiles of charge-transfer systems such as heterostructures of nickelates \cite{Fabbris2016, Wu2013, WrobelGeisler:18, Chen2020, GeislerPentcheva-LNOLAO-Resonances:19, Doennig2014, Geisler2020} and cuprates \cite{WrobelGeisler:18, GeislerPentcheva-NNOCCOSTO:21}.
Compared to these $e_g$ systems, orbital engineering is less explored in $t_{2g}$ systems, where, due to a weaker lattice coupling, the interplay between different degrees of freedom is more subtle \cite{Miyasaka2003, Geisler-Pentcheva2020}.
Here, we investigate these aspects in heterostructures of rare-earth vanadates ($R$VO$_3$).
We elucidate the relationship between the crystal and electronic structure of YVO$_3$ multilayers and thin films by combining experimental structural analysis with \textit{ab initio} calculations.

Rare-earth vanadate perovskites ($R$VO$_3$) are prototypical $t_{2g}$ systems that harbor a strong interplay of spin, orbital, and lattice degrees of freedom \cite{Horsch2008}. This gives rise to a rich phase diagram with multiple temperature-dependent phase transitions of spin ordering (SO) and orbital ordering (OO) patterns \cite{Horsch2008, Miyasaka2003, Sage2007}.
For example, YVO$_3$ exhibits three phase transitions,
below room temperature. At $\sim$200~K, a transition to $G$-type OO is observed, which is concomitant with a second-order structural phase transition.
This is followed by a $C$-type SO magnetic transition at $\sim$ 116~K. Finally, at $\sim$77~K, a first-order phase transition to a $C$-OO/$G$-SO phase occurs \cite{Miyasaka2006}.
The complex interactions in $R$VO$_3$ induce unusual phenomena such as temperature-dependent magnetization reversal \cite{Ren1998, Tung2007} and a spin-orbital entangled state \cite{Yan2019}. They have also attracted considerable attention as absorbing materials for high-efficiency solar cells \cite{Assmann2013, Brahlek2017}.
Over the past few decades, the nature of their ordering phenomena has been intensely debated in both theoretical \cite{Motome2003, Fang2004, Raychaudhury2007} and experimental studies \cite{Ulrich2003, Blake2009, Fujioka2010, Reul2012, Skoulatos2015}.
Despite a weaker Jahn-Teller interaction compared to $e_g$ systems, the spin and orbital properties are expected to be strongly coupled with the lattice, partly through the influence of lattice effects on the superexchange interactions \cite{Horsch2008, Yan2004}. This is also evident from the spin-orbital phase diagram of $R$VO$_3$ which varies widely with the V-O-V bond angle \cite{Miyasaka2003, Sage2007}.
Experimentally, this was examined in high-pressure studies, which showed that the ground state of $R$VO$_3$ can be altered under sufficient pressure \cite{Bizen2008, Zhou2009, Bizen2012}.
This lattice coupling was also inspected in thin films of LaVO$_3$ \cite{Vrejoiu2016, Meley2018, Vrejoiu2017, Meley2021} and PrVO$_3$  \cite{Kumar2019, Kumar2020}, where biaxial epitaxial strain was utilized to modify their ordering temperatures.

Recently, we explored the coupling of electronic and structural properties in YVO$_3$-LaAlO$_3$ (YVO-LAO) superlattices by combining x-ray resonant reflectometry with DFT+$U$ calculations \cite{Radhakrishnan2021}.
The results revealed that the bulklike degeneracy of the V-3$d$ $xz$ and $yz$ orbitals was lifted in all samples, independent of the YVO/LAO stacking sequences.
Further, within the YVO slabs, we found a reversal of the orbital polarization between the central and interface layers close to LAO. Temperature-dependent reflectometry measurements indicated that this phase was preserved down to 30~K. We qualitatively explained these results based on epitaxial strain and spatial confinement by LAO.

In this work, we delve deeper into the relation between the crystal structure and the orbital polarization patterns observed in Ref.~\cite{Radhakrishnan2021}, using scanning transmission electron microscopy (STEM) and x-ray diffraction (XRD), in conjunction with \textit{ab initio} calculations.
We also investigate the temperature-dependent crystal structure of a YVO film by using XRD, which suggests a suppression of bulklike structural phase transitions and provides a possible explanation for the absence of electronic transitions observed in related YVO multilayers \cite{Radhakrishnan2021}.
Additionally, we consider the effect of the substrate facet and the film thickness on the orbital polarization of the $xz$ and $yz$ orbitals.
The comparison of DFT+$U$ results for films versus superlattices indicates that the structural distortions and confinement effects induced by LAO in the multilayers, play an important role in enhancing the observed orbital polarizations.
We conclude that the substrate facet can be used to realize the desired film orientation for YVO films, which, together with the thickness of the YVO layers and the presence of spacer layers (such as LAO) in a multilayer, govern the resulting orbital polarization.

\section{METHODS}
\subsection{Experiments}
YVO-LAO superlattices (SLs) and YVO films of varying thicknesses were synthesized on the (110) facet of NdGaO$_3$(NGO) substrates, using pulsed laser deposition.
The SLs used in Ref.~\cite{Radhakrishnan2021} were grown with different stacking sequences of YVO/LAO: $(8/4)\!\times\!6$, $(6/6)\!\times\!6$, $(4/8)\!\times\!6$, pseudo-cubic unit cells~(u.c.). To study the thickness dependent properties, we additionally synthesized a 9 nm ultrathin YVO film and a thicker film of 23 nm thickness. To examine the YVO structure inside the multilayer, a 10 nm LAO film was also grown.
Finally, for investigating the facet dependence, YVO films were synthesized on (001) and (110) facets of NGO substrates having thicknesses of 42 nm and 44 nm, respectively.
Temperature dependent XRD was performed on a 49 nm YVO film on NGO(110) substrate.
An LAO capping layer of 4-5 nm thickness was grown on all films to preserve their oxidation state.
All samples were grown with the parameters listed in Ref.~\cite{Radhakrishnan2021}. Structural analysis was done using XRD and STEM.
A JEOL JEM-ARM 200F scanning transmission electron microscope was used to perform the STEM measurements. The simulated high-angle annular dark-field (HAADF) and annular bright-field (ABF) images were generated using the multi-slice method implemented in the QSTEM image simulation software.
The optical parameters used for the simulation were the same as the experimental values.
The thickness used in the STEM image simulation was 20 nm.
Room temperature XRD measurements (in Sec.~\ref{sec-A} and \ref{sec-C}) were performed using in-house laboratory diffraction set-up and temperature dependent measurements  (in Sec.~\ref{sec-B}) were performed with the closed-circle cryostat set-up at KARA, MPI beamline, Karlsruhe. Lattice parameters were refined using the software CELREF~\cite{celref}.
The XRD scans are displayed in units of 1/$d$ and momentum transfer $Q$, with $d$ = $\lambda$/2$\sin \theta$ and $Q$ = 2$\pi$/$d$, where $\theta$ and $\lambda$ are the Bragg angle and wavelength of x-rays, respectively.
Throughout this article, the orthorhombic $Pbnm$ (no. 62) space group is used to index reflections and directions.
The subscript $\textit{pc}$ is used to indicate reflections where a pseudocubic notation is used.

\subsection{Theory}
We performed first-principles calculations in the framework of spin-polarized density functional theory~\cite{KoSh65} (DFT) as implemented in the Quantum ESPRESSO code~\cite{PWSCF}.
The generalized gradient approximation was used for the exchange-correlation functional as parametrized by Perdew, Burke, and Ernzerhof~\cite{PeBu96}. Static correlation effects were considered within the DFT$+U$ formalism~\cite{QE-LDA-U:05} ($U_\text{V}=3$~eV).
To account for octahedral rotations, orbital order, and magnetic order simultaneously, we model the YVO-LAO SLs by using large $p(2 \times 2)$ (4/4) supercells that are consistent with the experimental realization of four pseudocubic unit cells per layer.
In addition, we explicitly treat interfacial Y-La intermixing. Pure YVO films were modeled as strained bulk in a comparable $p(2 \times 2)$ supercell, which contains eight distinct V sites in total.
In all cases, the atomic positions were fully optimized.

\begin{figure*}[t]
	\includegraphics[width=1.9\columnwidth]{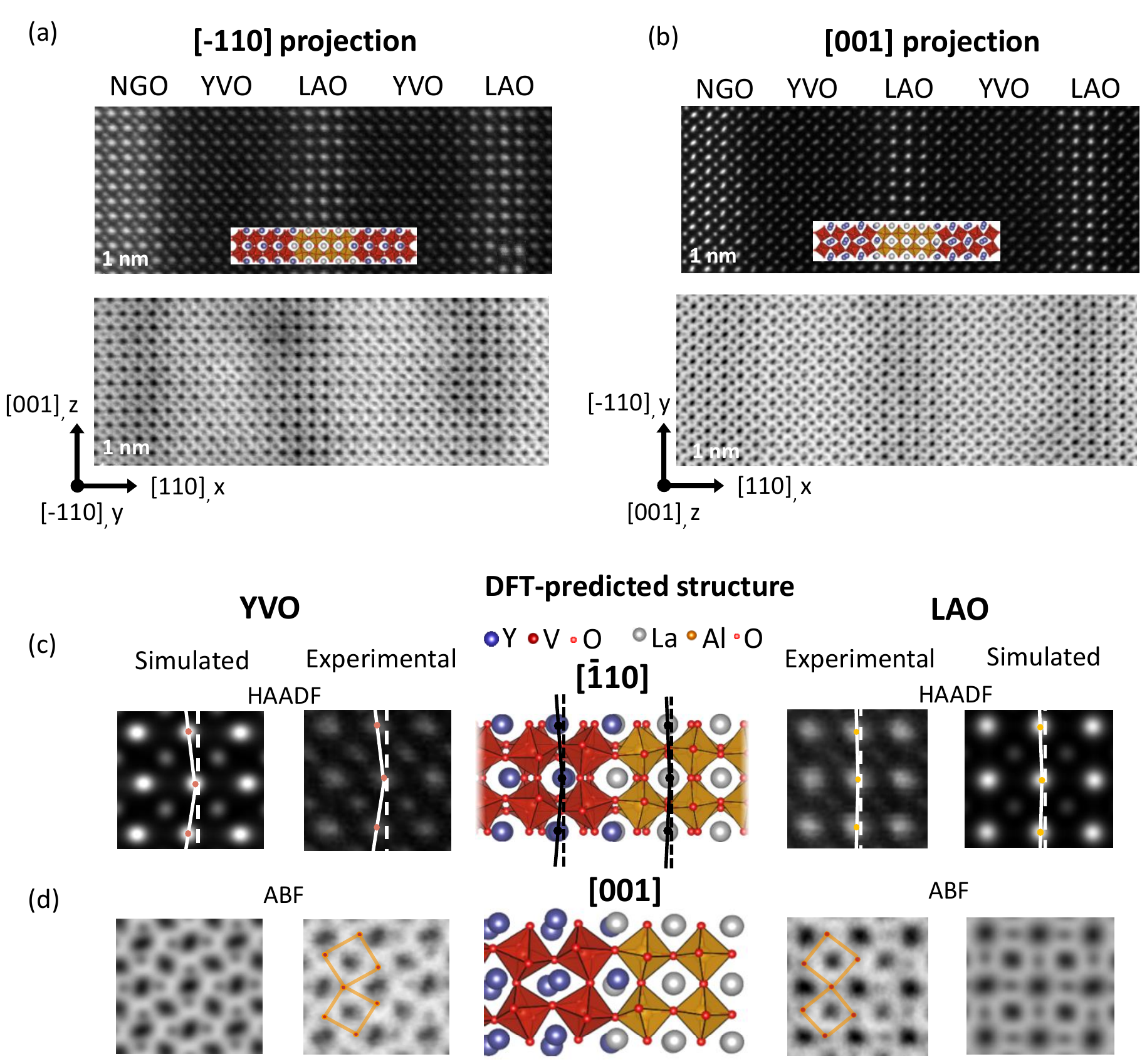}
\caption{HAADF (top) and ABF (bottom) images normal to the (a) $[-110]$ and (b) $[001]$ directions for the YVO/LAO: (8/4) SL, together with the optimized DFT+$U$ SL structure. (c, d) Simulated and experimental $[-110]$-projected HAADF and $[001]$-projected ABF images, respectively, of the YVO (left) and LAO regions (right) in the SL. The optimized DFT+$U$ interface geometry shown in the center, was used to simulate the HAADF images, which closely agree with experiment. Red and yellow octahedra correspond to YVO and LAO in the SL, respectively.}
\label{fig1}
\end{figure*}

\section{RESULTS AND DISCUSSION}
\subsection{Prediction and verification of crystal structure}
\label{sec-A}
\label{discussion}
We start our discussion with the crystal structure of the SLs and demonstrate that DFT+$U$ calculations are able to accurately predict even their minutest structural aspects.
A detailed examination of the STEM images confirms that the SLs indeed exhibit the octahedral rotation patterns predicted by DFT+$U$.
Further, we also verify the bond-angle trends across the interface from the calculations by an extensive XRD analysis of the SLs and thin films.

In Fig.~\ref{fig1} (a) and (b), HAADF and ABF STEM images of the (8/4) SL are shown, which were taken along its two cross-sectional directions.
The absence of defects and stacking faults in the low and high magnification images (Fig.~\ref{fig0} and \ref{fig1}), indicates the excellent sample quality,
on a wide length scale from a few angstroms to micrometers.
The orientation of YVO and LAO unit cells in the SLs was determined by using XRD and was found to follow the $(110)$ facet of the NGO substrate, with the distinct orthorhombic-type \cite{ft2} $c$-axis being in-plane and parallel to that of NGO \cite{Radhakrishnan2021}.
Thus, the cross-sectional images across the two in-plane directions of the sample yield the $[001]$ and $[-110]$ projections.
While HAADF images provide better $Z$-contrast, i.e., heavier cations with large atomic number $Z$ (here Y and La) appear brightest, the contrast is inverted in ABF images and therefore the positions of lighter elements such as oxygen can be mapped with higher precision.
In the $[001]$ projection, the oxygen atoms along the transmission direction, fall on top of each other, creating a single column, and therefore the oxygen positions are resolvable in this projection.
On the other hand, viewed along the $[-110]$ direction, the Y and La cations are co-incident with subsequent layers, making their positions resolvable in this projection.
Thus, ABF in the $[001]$ and HAADF in the $[-110]$ projections are used for resolving oxygen and cation positions, respectively.

Fig.~\ref{fig1}(c) and (d) provide a detailed analysis of the YVO-LAO interface region obtained from STEM and DFT+$U$, for the HAADF and ABF images, respectively.
The experimental images are magnifications of the HAADF and ABF images [from Fig.~\ref{fig1}(a, b)] and the simulated images were generated using the DFT+$U$-predicted structure, shown in the middle.
We first focus on the HAADF image of the YVO region, where the clearly resolved Y cations create a characteristic zig-zag pattern along the $c$-axis (parallel to $z$) [Fig.~\ref{fig1}(c), left]. These observations from STEM match closely with the optimized DFT+$U$-geometry as well as the STEM simulation. They reveal that YVO has an a$^-$b$^-$c$^+$ Glazer tilt system \cite{Glazer1975} in the SL, similar to bulk YVO, which has an a$^-$a$^-$c$^+$ tilt system \cite{Martinez2008}. The a, b, and c tilts are rotations about the $x$, $y$, and $z$ pseudocubic directions, respectively, for the reference frame displayed in Fig.~\ref{fig1}(a, b).
\begin{figure*}[tb]
	\includegraphics[width=1.7\columnwidth]{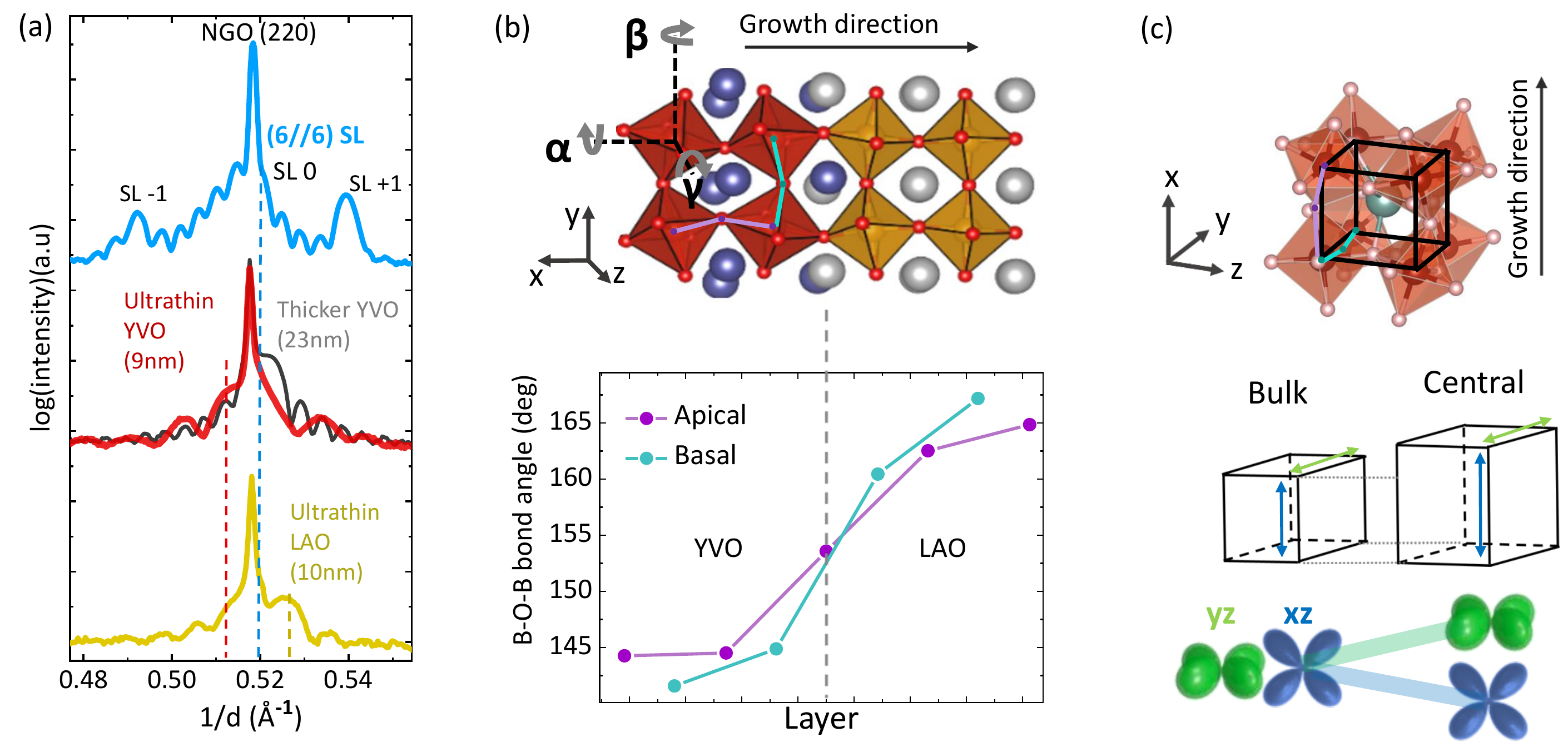}
\caption{(a) XRD of out-of-plane (220) reflections of the (6/6) SL shown in blue (top), ultrathin YVO film (9 nm) and the thicker YVO film (23 nm) shown in red and grey respectively (middle), ultrathin LAO film (10 nm) shown in yellow (bottom).
(b) Optimized DFT+$U$ SL structure (top) and in-plane and out-of-plane bond angles as predicted by DFT calculations (bottom). The apical and basal $B$-O-$B$ angles are marked in purple and turquoise, respectively.
(c) Pseudocubic YVO unit cell (top panel) displaying the apical and basal bond angles with respect to schematic (bottom panel) showing the lifting of degeneracy in the C layers, due to an increase in the out-of-plane lattice parameter.}
\label{fig2}
\end{figure*}
We assign different magnitudes (a, b, c) to each tilt in order to consider the most general case for the YVO in the SL \cite{Glazer1975}, since the magnitude of the tilts cannot directly be discerned from the projected STEM images. We will later discuss the tilt pattern of YVO in the context of other studies (Sec.~\ref{sec-B}).

Now we turn to the LAO layers, which in the bulk, feature a rhombohedral symmetry with a$^-$a$^-$a$^-$ tilts \cite{Lehnert2000}.
As expected, the octahedral tilts in LAO are much smaller than those of YVO, as observed from the ABF images [Fig.~\ref{fig1}(d)].
Interestingly, DFT predicts that in the SL geometry, LAO will follow the tilt system of YVO and adopt an a$^-$a$^-$c$^+$-type \cite{ft2} pattern.
In the HAADF image of LAO [Fig.~\ref{fig1}(c), right], we see a clear signature of a plus tilt, i.e., the cation displacements characteristic of the plus tilt are visible in LAO, albeit much more subtly, compared to those of YVO.
Since, these displacements are absent in the rhombohedral system, we conclude that the LAO region in the SL possesses orthorhombic-type symmetry, in line with the DFT predictions.
In this regard, we note that other aluminate perovskites such as YAlO$_3$ possess a $Pbnm$ structure with a$^-$a$^-$c$^+$ tilts, indicating the close-lying energies of both structures.
Furthermore, this observation exemplifies the importance of octahedral connectivity across interfaces for the resulting structure stabilization in ultrathin layers.
Since LAO merely acts as a confinement layer in the present system without an electronically active role, the change in its tilt pattern has no direct consequence on the electronic structure of the SL.
However, this demonstrates that the DFT$+U$ calculations are able to predict the specific crystal structure of each component within the SL with remarkable accuracy.
We will exemplify this further when we compare the results of XRD with the DFT predictions of bond angles.

We complement the electron microscopy results of the YVO region in the SL by XRD investigations of individual LAO and YVO films that provide sufficient sample volume.
Fig.~\ref{fig2}(a) displays the XRD of out-of-plane reflections of an ultrathin YVO film ($\sim$9 nm) and a thicker film ($\sim$23 nm), indicating different out-of-plane lattice parameters of the two films.
Note that the corresponding Bragg reflections from a SL is an average of the individual constituents of the multilayer.
Hence, in order to produce the average position of the SL Bragg peak, the structure of the YVO inside the SL must be similar to the ultrathin film.
Despite the tensile strain, the ultrathin film has a larger lattice parameter than its bulk counterpart, implying a negative Poisson ratio, which can arise from the necessity to maintain octahedral connectivity, as was reported in Ref.~\cite{J.Hwang2013}.
We explain this as follows: for simplicity, we consider rigid octahedra and use the pseudocubic axes: $x$, $y$, $z$ to define the frame of reference for the rotation of the octahedra [Fig.~\ref{fig2}(b)].
As was shown by A.M.~Glazer \cite{Glazer1975}, for $AB$O$_3$ perovskites, rotation \textit{about} an axis ($x$, $y$ or $z$) changes the lattice parameters in the two directions perpendicular to it. Alternatively, tilting \textit{along} a certain axis changes the lattice parameter along that axis.
YVO is mostly under tensile strain along its $c$-axis, thus as the cations along $z$ move farther apart, the tilt along this direction would be reduced. To accomplish this, the octahedra have two options: either to rotate about $x$ or to rotate about $y$. Rotating about $x$ will require in-plane cation movement, but since the in-plane lattice parameters are fixed by the substrate, the rotation is likely to happen about $y$, increasing the out-of-plane lattice parameter in the process, in agreement with our observations.
We will later return to this aspect and highlight the importance of octahedral connectivity in the present system.

Next, we inspect the evolution of the $B$-O-$B$ bond angles across the YVO-LAO interface as predicted by DFT$+U$ [Fig.~\ref{fig2}(b)].
The apical (out-of-plane) and basal (in-plane) values refer to the bond angles along $x$ and $y$ directions, respectively.
Comparing the two, the apical angles are larger (which corresponds to a less pronounced rotation) than the basal ones for YVO.
This promotes a larger out-of-plane lattice parameter, as we indeed observed from XRD.
Fig.~\ref{fig2}(c) illustrates the effect of this structural deformation on the electronic structure of the SLs:
YVO in the SL has a larger out-of-plane parameter (along $x$) compared to bulk YVO, but the in-plane parameter (along $y$) is close to the bulk value.
Due to the antibonding nature of the V $3d$ orbitals, a larger lattice parameter lowers the energy of the orbitals with lobes along that direction, and thus lifts the degeneracy between the $xz$ and $yz$ orbitals \cite{Rogge2018}.
Here, we recognise another crucial point regarding the orientation of the YVO unit cell.
It is clear from Fig.~\ref{fig2}(c), that the degeneracy between these orbitals would only be removed if the $c$-axis (parallel to $z$) of YVO lies in the plane of the sample.
This results from the fact that both strain and confinement effects operate by differentiating between the in-plane and out-of-plane directions.
We will further explore the influence of the out-of-plane parameter and the unit-cell orientation in the discussion of our \textit{ab initio} calculations in Sec.~\ref{sec-D}.

\subsection{Temperature-dependent x-ray diffraction}
\label{sec-B}
The lack of a temperature dependence of the resonant reflectometry data \cite{Radhakrishnan2021} is a clear indication of the suppression of bulklike phase transitions in the YVO SL.
To investigate the structural modifications induced by the heteroepitaxy as a possible cause of this suppression, we performed temperature-dependent XRD on a YVO thin film, ranging from 40\,K to 270\,K.
We chose a film for this purpose instead of the SL, due to the proximity of the SL Bragg peak and substrate peaks for many reflections, which imposed challenges in determining their positions.
\begin{figure}[tb]
\center\includegraphics[width=\columnwidth]{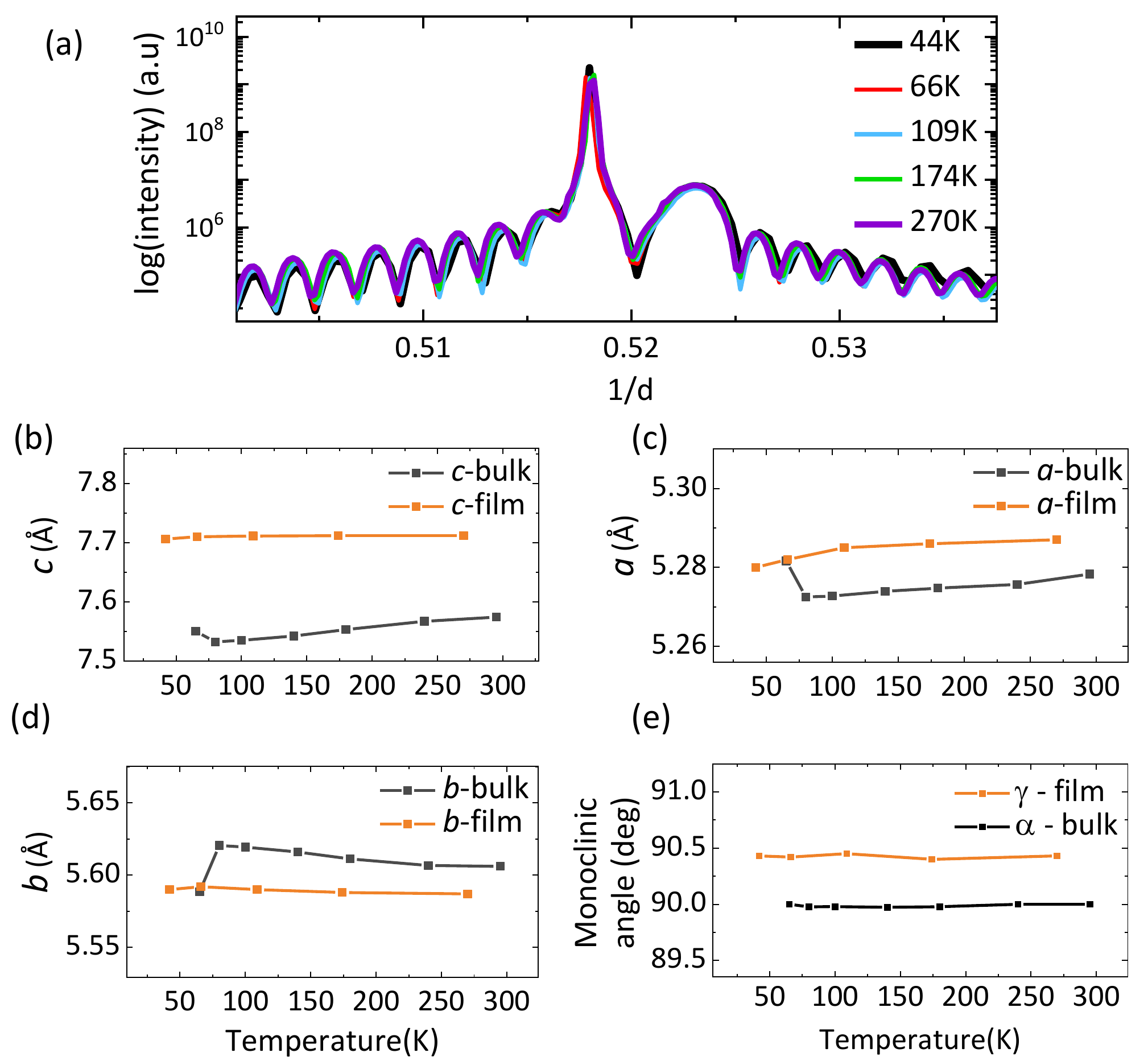}
\caption{Temperature dependence of YVO film on NGO(110) substrate; (a) Out-of-plane XRD scan of (220) reflection.
(b-e) Temperature dependence of lattice parameters: $a$, $b$, $c$, and the monoclinic angle of film in $P2_1/m$ symmetry (no.11), with respect to bulk YVO in $P2_1/b$ symmetry (no.14) (taken from \cite{Blake2002, ft}).}
\label{fig3}
\end{figure}

\begin{figure*}[tb]
\center\includegraphics[width=1.9\columnwidth]{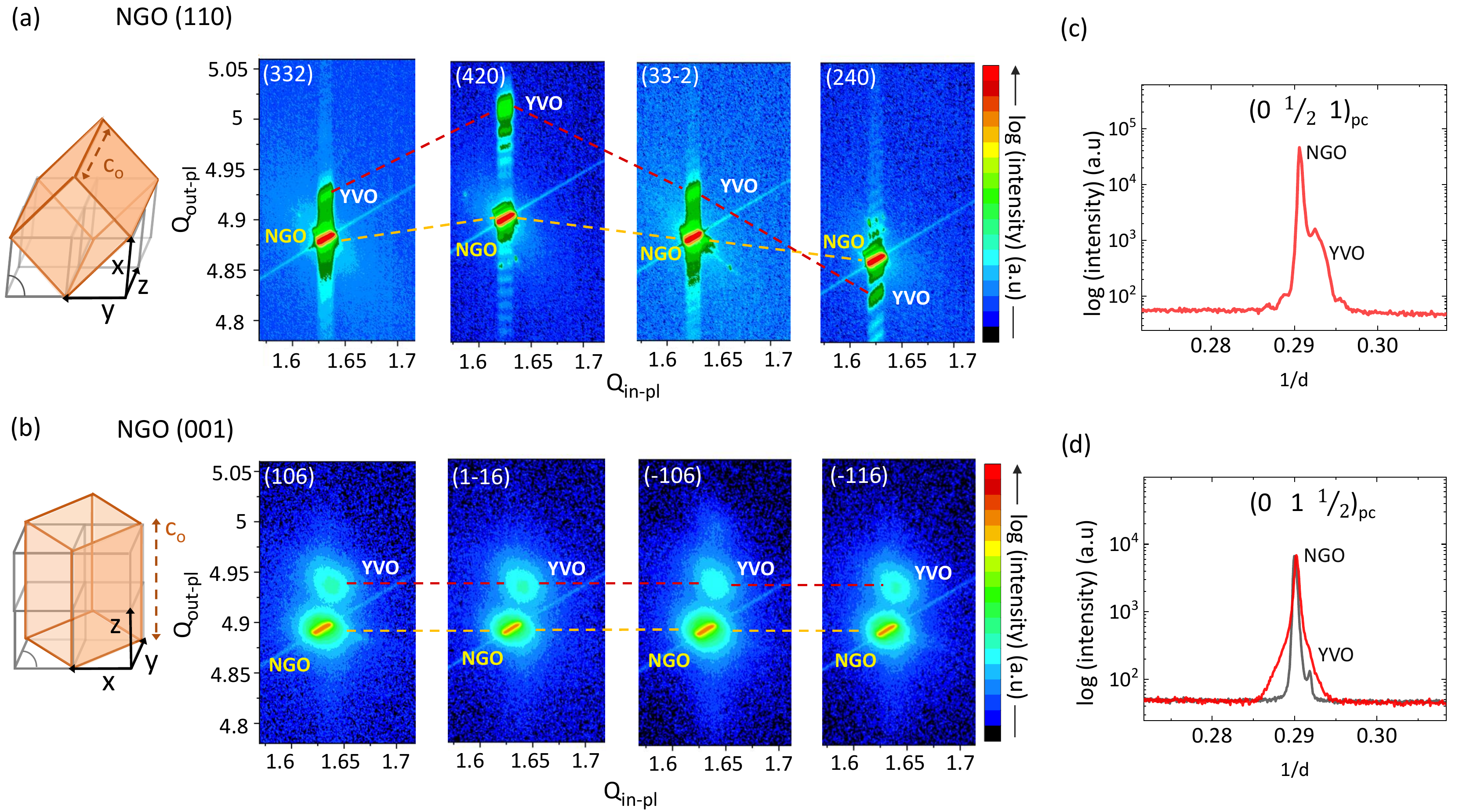}
\caption{(a,b) Reciprocal space maps of the $\{$103$\}_{\text{pc}}$ family of reflections of YVO thin films grown on NGO(110) and NGO(001) substrates, respectively.
The corresponding orthorhombic Miller indices are indicated for each map. The red and yellow dotted lines trace the evolution of peak positions for the films and substrates, respectively. The schematics display the orientation of the orthorhombic-type film unit cell (orange) superimposed on the corresponding pseudocubic unit cells (grey).
(c,d) Half-order reflections (0 \nicefrac{1}{2} 1)$_\text{pc}$ and (0 1 \nicefrac{1}{2})$_\text{pc}$, shown in red, for the films on NGO(110) and NGO(001) facets, respectively. The substrate peak for the (0 1 \nicefrac{1}{2})$_\text{pc}$ reflection is shown in grey, in part (d).}
\label{fig4}
\end{figure*}
The film lattice parameters were calculated using the room temperature substrate peak positions as a reference, by translating all substrate peaks through temperature to the room temperature value and neglecting the effects of thermal expansion of the substrate.
The temperature dependence of the (220) reflection of the film [Fig.~\ref{fig3}(a)] suggests that the out-of-plane lattice parameter of YVO shows negligible variation with temperature.
To extract the lattice parameters of the film, we utilized seven Bragg reflections: (332), (420), (150), (334), (510), (442) and (530).
Using the software \textsc{CELREF} \cite{celref}, the lattice parameters were refined in the monoclinic $P2_1/m$ symmetry, i.e., for parameters $a$, $b$, $c$, and $\gamma$.
The temperature dependence of lattice parameters of the film and bulk YVO (reproduced from \cite{Blake2002}) are shown in Fig.~\ref{fig3} (b)-(e).

At room temperature, the largest deviation, compared to the bulk, is observed in the in-plane $c$ parameter [Fig.~\ref{fig3}(b)], arising from the fact that the tensile strain exerted by the substrate is largest along this direction.
Due to the (110) orientation of the film, the deformations of $a$, $b$, and the angle $\gamma$ between the two, are coupled. Here, the film becomes monoclinic to accommodate the strain by increasing $\gamma$ above 90$^{\circ}$ and reducing the difference between the $a$ and $b$ parameters, i.e., the orthorhombicity ($b$/$a$) of the system \cite{Vailionis2011}. This behaviour is reminiscent of other orthorhombic systems under tensile strain, which were identified to have the a$^-$b$^-$c$^+$ tilt pattern in the space group of $P 1 1 2_1/m$ (no. 11) \cite{Vailionis2011}.
This is a different monoclinic sub-group than that of bulk YVO in the intermediate temperature range (77~K$<T<$200~K), which exhibits the $P2_1/b 1 1$ (no.\ 14) space group, with $\alpha$ (between $b$ and $c$) being the monoclinic angle \cite{Reehuis2006}. Thus, in Fig.~\ref{fig3}(e), we show the monoclinic angles of both the film and bulk YVO.

In contrast to the lattice parameters of bulk YVO, the temperature dependence of film lattice parameters shows almost no changes [Fig.~\ref{fig3}(b-e)].
The bulklike first-order structural transition at 77~K is suppressed in the YVO film. The second-order structural transition at 200~K is more subtle and thus difficult to observe from the trends of the lattice parameters.
We assume that the absence of the first-order transition, verified here for the film, can be extended to the SLs as well, since they are also strained to the substrate.
In particular, the presence of the in-plane orientation of the $c$-axis appears to be a common, important prerequisite for the suppression of structural transitions, as was suggested in Ref.~\onlinecite{Meley2018}, for epitaxial LaVO$_3$ thin films.

Calculating the corresponding pseudocubic parameters, as described in Ref.~\cite{Vailionis2011}, we obtain $a_\text{pc} = 3.85$\AA, $b_\text{pc} = 3.86$\AA, $c_\text{pc} = 3.83$\AA\, which are nearly constant through temperature. The in-plane parameters ($a_\text{pc}$ and $b_\text{pc}$) match those of the substrate, due to being epitaxially strained, which serves as a validation of the refinement process.
Since in bulk YVO, the orbital ordering phase transitions are concomitant with structural transitions, their absence in the film is expected to also affect its electronic structure.
This result fits well with the fact that the electronic structure obtained from reflectometry also did not show any temperature-dependent changes from room temperature to 30~K \cite{Radhakrishnan2021}.

\subsection{Impact of the growth facet}
\label{sec-C}
As mentioned earlier, in Sec.~\ref{sec-A}, we expect the film orientation to have a profound impact on its electronic structure.
Therefore, in this section, we examine the effect of the substrate facet on the crystal structure of YVO films grown on (110) and (001) facets of NGO substrates, using XRD.
In the next section (Sec.~\ref{sec-D}), we will then inspect their electronic structure.

Fig.~\ref{fig4}(a, b) shows the reciprocal space maps of the $\{$103$\}_\text{pc}$ family of reflections of YVO films grown on the (110) and (001) facets of NGO, respectively.
These reflections are measured at four azimuthal angles ($\phi$), that are 90$^{\circ}$ apart.
For the (110) facet [Fig.~\ref{fig4}(a)], the evolution of position of peaks as a function of the azimuthal angle (dotted lines) is the same for both, the substrate and the film.
This characteristic pattern implies that the pseudocubic unit cell is tilted with respect to the azimuth ($z$) \cite{Vailionis2011}. This means that the orthorhombic-type unit cell for both film and substrate is oriented as shown in the schematic, with their $c$-axes in the plane of the substrate surface, and also parallel to one another.
In contrast to this, for the (001) facet [Fig.~\ref{fig4}(b)], both the substrate and film peaks fall in one horizontal line through the variation of $\phi$. This indicates that here, the $c$-axes of the film and the substrate both point out of the plane, perpendicular to the substrate surface, as shown in the schematic.

To further confirm these inferences, we measured half-order reflections, that directly probe the cation displacements associated with the orthorhombic-type space groups.
Due to the doubling of the unit cell along the $c$-axis, half-order reflections that are forbidden for the undistorted perovskite structure are allowed for the orthorhombic-type structure.
When indexed in the pseudocubic unit cell, the position of the fractional Miller index (in $h$, $k$, or $l$) indicates the direction along which the unit cell doubles, i.e., the direction of the $c$-axis.
Thus, the presence of (\nicefrac{1}{2} 0 1), \mbox{(0 \nicefrac{1}{2} 1)} reflections and (0 1 \nicefrac{1}{2}), (1 0 \nicefrac{1}{2}) reflections imply that the $c$-axis lies in-plane and out-of-plane of the sample, respectively \cite{Meley2018}.

As expected for (110) orientation, we only found an intense (0 \nicefrac{1}{2} 1) reflection for the substrate and film [Fig.~\ref{fig4}(c)], with other peaks being absent.
On the other hand, for the (001) facet [Fig.~\ref{fig4}(d)], we found a strong (0 1 \nicefrac{1}{2}) reflection for both the substrate and film. Peaks for the (\nicefrac{1}{2} 0 1) and \mbox{(0 \nicefrac{1}{2} 1)} were very weak and absent respectively, indicating the presence of a negligible percentage of twin domains. Since the film peak occurs at almost the same position as the substrate, we verified the presence of the film by performing the same scan for a bare NGO(001) substrate. These results confirm that the choice of the substrate facet can be used to obtain the desired orientation of the YVO film, since the film adopts the same orientation as that provided by the substrate surface.

\subsection{Insights from first-principles simulations}
\label{sec-D}

\begin{figure}[tb]
\center\includegraphics[width=8.1cm]{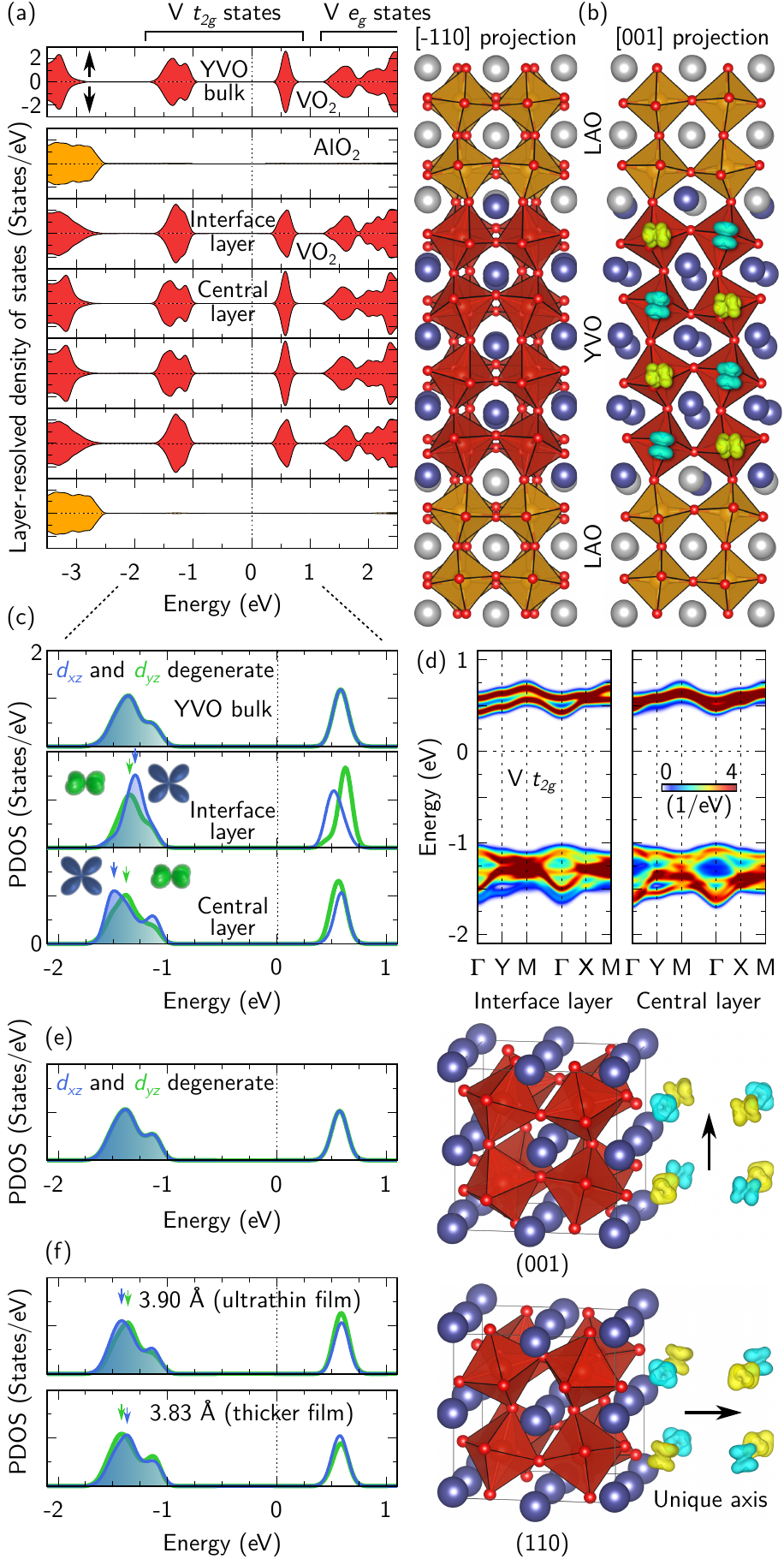}
\caption{Electronic structure (a) and optimized geometry (b) of the representative YVO/LAO 4/4 SL. Spin densities in yellow and turquoise highlight the $G$-type ($C$-type) AFM SO (OO). (c) PDOS of the V-3$d$ $xz$ and $yz$ orbitals. (d)~Corresponding SL band structure, highlighting contributions of the V-$t_{2g}$ manifold. (e,f) Results for YVO, strained to NGO(001) and NGO(110) substrates. The different out-of-plane lattice parameters for the (110) case mimic the distinct experimental values observed for ultrathin versus thicker YVO films, inducing different orbital energy sequences. }
\label{fig5}
\end{figure}
We have demonstrated above and in Ref.~\cite{Radhakrishnan2021} that DFT$+U$ accurately describes both the crystal geometry and the electronic structure of the YVO-LAO SLs.
Now, we discuss first-principles results that address the influence of the thickness and substrate growth facet on the electronic structure of YVO films,
and compare these results with the YVO electronic structure in the SL geometry with LAO.

For the YVO film on the NGO(001) substrate, we find that it is energetically favorable by $5.5$~meV per V-ion to align the distinct $c$-axis with the out-of-plane direction as compared to placing it in the plane.
This agrees with the experimental observations discussed above (see Fig.~\ref{fig4}).
Interestingly, even on the NGO(110) substrate, such an alignment is still energetically favourable, though by a smaller amount of $2.6$~meV/V-ion, owing to a slightly better accommodation of the epitaxial strain.
However, this would cause a discontinuity in the octahedral rotation pattern at the YVO/NGO interface.
We therefore conclude that the experimentally observed alignment of the $c$-axis parallel to that of the substrate
is a consequence of the octahedral connectivity,
which highlights its important role in these strongly tilted oxide systems.
For comparison, typical energy differences for quenching or altering the a$^-$a$^-$c$^+$ octahedral rotation pattern in YVO bulk are of the order of several hundreds of meV per V-ion.
Moreover, canceling the octahedral rotations in the less tilted LaNiO$_3$/LAO(001) SLs requires $250$-$300$~meV per transition metal ion~\cite{GeislerPentcheva-LNOLAO:18, GeislerPentcheva-LNOLAO-Resonances:19}.

Fig.~\ref{fig5}(a) shows the layer-resolved electronic structure of the representative 4/4 SL as obtained from DFT$+U$ simulations.
The V-$3d$ states are located deep within the band gap of the insulating LAO, as the valence band maximum of the latter is located at about $-2.5$~eV.
The V-$t_{2g}$ states can be observed between $-1.7$ and $-1.0$~eV and at $0.6$~eV, while the V~$e_{g}$ states are located at higher energies $> 1.2$~eV.
At first glance, the V electronic structure in the SL resembles that in YVO bulk, with the $xy$ orbital occupied throughout, and an alternating occupation of the $xz$ and $yz$ orbitals.
This $C$-type OO, which is accompanied by a $G$-type AFM SO, is illustrated by the spin density in Fig.~\ref{fig5}(b),
superimposing the optimized SL structure, which is shown along two directions perpendicular to the growth direction.
A closer inspection of the projected density of states (PDOS) [Fig.~\ref{fig5}(c)] reveals that the $xz,yz$ degeneracy of YVO bulk is lifted in the heterostructure,
and in the opposite manner in the interface layers with respect to the central layers.
This observation has already been emphasized in our previous work \cite{Radhakrishnan2021},
which confirmed experimentally that the orbital polarization between the $xz$ and $yz$ orbitals is inverted for the central layers with respect to the interface layers,
whereas the $xy$ orbital continues to have the largest occupation for both layers (as in YVO bulk).
The distinction between interfacial and central V sites is also reflected by the $k$-resolved densities of states (band character plot) [Fig.~\ref{fig5}(d)].

Our findings can be compared to the electronic structure of YVO when strained to different substrate facets,
which describes the situation of the two epitaxially grown films on different NGO facets.
For films grown on NGO(001), where the $c$-axis (coinciding with the unique axis) is out-of-plane, the calculations show that
the V-3$d$  $xz$ and $yz$ orbitals retain their bulklike degeneracy [Fig.~\ref{fig5}(e)].
Furthermore, we do not observe substantial changes in the splitting between the occupied and the unoccupied V-$t_{2g}$ states.
On the other hand, for films grown on NGO(110), where the $c$-axis is in the plane, we clearly see that the $xz,yz$ degeneracy is lifted [Fig.~\ref{fig5}(f)].
The energy sequence of the orbitals is reversed between the ultrathin and thicker YVO films and is determined by their out-of-plane lattice parameters of 3.90 and 3.83 \AA, respectively, which were obtained from XRD.
Interestingly, the solitary impact of epitaxial strain on the V-$t_{2g}$ orbitals, illustrated in Fig.~\ref{fig5}(f),
is smaller than the concerted effect of strain, quantum confinement, and octahedral connectivity in the SL geometry with LAO [Fig.~\ref{fig5}(c)].
These results confirm the conclusions made in our previous study \cite{Radhakrishnan2021}, i.e. the finite orbital polarization arising in the SL geometry is not only substantially modified by structural distortions, but also by the electronic confinement by LAO.
Moreover, our DFT+$U$ calculations disentangle the impact of these aspects on the YVO electronic structure from that of epitaxial strain.
We see that the substrate growth facet has a profound effect on the electronic structure of YVO films.

\section{CONCLUSION}
In the present study, we investigated the relation between the crystal and electronic structure of YVO-LAO SLs and YVO epitaxial thin films.
First-principles simulations accurately predict the crystal structure modifications that lead to the lifting of degeneracy between the V-3$d$ $xz$ and $yz$ orbitals, observed in Ref.~\cite{Radhakrishnan2021}. These predictions are in good agreement with the experimental results of XRD and STEM. Using DFT+$U$ simulations, we examined the influence of the YVO layer thickness and the orientation of its unit cell on the orbital polarization.
The results show that the finite orbital polarization occurs only if the $c$ axis is oriented in-plane.
The order of preferred net occupation of $xz$ and $yz$ orbitals depends sensitively on the out-of-plane lattice spacing and even shows a reversal between 3.83 and 3.90\AA .
The quantitative comparison of the calculations for YVO under epitaxial strain versus YVO-LAO SLs reveals an enhancement of orbital polarization in the latter, implying that the local structural changes and confinement effects created by LAO in the superlattice geometry, play an important role in the electronic reconstruction.
Finally, temperature-dependent XRD suggested that bulklike structural phase transitions of YVO are suppressed in the film, which agrees with the lack of temperature dependence of the electronic structure observed for the SLs \cite{Radhakrishnan2021}.

The close agreement between the detailed structural analysis and DFT$+U$ calculations demonstrates how the
latter can reliably predict the electronic reconstructions in complex oxide heterostructures.
In particular, our findings on the influence of the substrate facets and concerted effects of structural modifications and electronic confinement at interfaces are of general relevance for many perovskite heterostructures with $Pbnm$ crystal symmetry.
Our study underlines the key role of structural modifications on electronic properties, and illustrates the use of orbital engineering as a promising approach for the theory guided rational design of correlated materials.

\begin{acknowledgments}
This work was supported by the German Research Foundation (Deutsche Forschungsgemeinschaft, DFG) within the SFB/TRR~80 (Projektnummer 107745057), Project No.~G1 and G3. Computing time was granted by the Center for Computational Sciences and Simulation of the University of Duisburg-Essen (DFG Grants No.~INST 20876/209-1 FUGG and No.~INST 20876/243-1 FUGG). This project has also received funding from the European Union's Horizon 2020 research and innovation program under grant agreement No. 823717 - ESTEEM3. The KIT Institute for Beam Physics and Technology (IBPT) is acknowledged for the operation of the storage ring, the Karlsruhe Research Accelerator (KARA), and provision of beamtime at the KIT light source.
\end{acknowledgments}


\begin{thebibliography}{68}
\bibitem [{ft2()}]{ft2}%
  \BibitemOpen
  \href@noop {} {}\bibinfo {note} {We use the term `orthorhombic-type' to
  collectively refer to $Pbnm$ (a$^-$a$^-$c$^+$ tilts) and its monoclinic
  sub-space groups with a$^-$b$^-$c$^+$ tilt pattern (see
  Sec.~\ref{sec-B}).}\BibitemShut {Stop}%
\end{thebibliography}
\end{document}